\documentclass{article}

\usepackage{arxiv}
\usepackage[utf8]{inputenc} % allow utf-8 input
\usepackage[T1]{fontenc}    % use 8-bit T1 fonts
\usepackage{hyperref}       % hyperlinks
\usepackage{url}            % simple URL typesetting
\usepackage{booktabs}       % professional-quality tables
\usepackage{amsfonts}       % blackboard math symbols
\usepackage{nicefrac}       % compact symbols for 1/2, etc.
\usepackage{microtype}      % microtypography
\usepackage{lipsum}		% Can be removed after putting your text content
\usepackage{graphicx}
\usepackage{natbib}
\usepackage{doi}
\usepackage{caption}
\usepackage{subcaption}
\usepackage{algorithm}
\usepackage{algorithmic}
\usepackage{xcolor}
\usepackage{varwidth}
\usepackage[english]{babel}
\usepackage{booktabs}
\usepackage{multirow}
\usepackage{amsmath}
\usepackage{cleveref}
\usepackage{float} % allows [ht] specifier

\definecolor{HH}{RGB}{255,255,179}
\definecolor{HC}{RGB}{251,128,124}
\usepackage[normalem]{ulem}
\newcommand{\smallorcid}{%
  \smash{\raisebox{-0.3ex}{\includegraphics[scale=0.06]{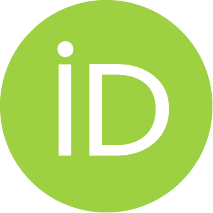}}}%
}

\date{} 	% Or removing it

\title{Hybrid Intelligence in Cartoon Captioning: Evaluating AI as a Creative Writing Partner}

\author{
\resizebox{\textwidth}{!}{
\begin{tabular}{ccccc}
\href{https://orcid.org/0000-0001-8347-3038}{\smallorcid\hspace{1mm}U\u{g}ur \"{O}nal}$^{1}$ &
\href{https://orcid.org/0000-0003-2993-6681}{\smallorcid\hspace{1mm}Sanem Sariel}$^{2}$ &
\href{https://orcid.org/0000-0002-1524-1646}{\smallorcid\hspace{1mm}Metin Sezgin}$^{3}$ &
\href{https://orcid.org/0009-0000-4712-1385}{\smallorcid\hspace{1mm}Derya Akleman}$^{4}$ &
\href{https://orcid.org/0000-0003-3618-4166}{\smallorcid\hspace{1mm}Ergun Akleman}$^{1,5}$
\\
\texttt{uguronal@tamu.edu} &
\texttt{sariel@itu.edu.tr} &
\texttt{mtsezgin@ku.edu.tr} &
\texttt{akleman@tamu.edu} &
\texttt{ergun@tamu.edu} 
\end{tabular}
}
\\[1em]
{
\begin{tabular}[t]{@{}c@{}}
$^{1}$College of Performance, Visualization \& Fine Arts, Texas A\&M University, College Station, TX, 77843\\
$^{2}$Artificial Intelligence and Data Engineering Dept., Istanbul Technical University, Turkey\\
$^{3}$Computer Science and Engineering Dept., Koc University, Istanbul, Turkey\\
$^{4}$Department of Statistics, Texas A\&M University, College Station, TX, 77843\\
$^{5}$Joint with Computer Science and Engineering Dept., Texas A\&M University, College Station, TX, 77843
\end{tabular}
}
}

\renewcommand{\shorttitle}{Hybrid Humor}

\hypersetup{
  pdftitle={Hybrid Humor: Investigating AI’s Potential in Cartoon Caption Writing},
  pdfsubject={Artificial intelligence; computational creativity; humor; cartoon captioning; human--AI collaboration; user study},
  pdfauthor={U\u{g}ur \"{O}nal; Sanem Sariel; Metin Sezgin; Derya Akleman; Ergun Akleman},
  pdfkeywords={cartoon captioning, humor generation, large language models, ChatGPT, GPT-4o, human-AI collaboration, computational creativity, visual storytelling, user study, authorship perception},
}

\renewcommand{\undertitle}{}

\begin{document}

\vspace*{-3em}
\maketitle
\vspace*{-4em}

\begin{figure}[ht]  % H = put it exactly here
    \centering
    \begin{subfigure}[b]{0.495\textwidth}
        \includegraphics[width=\textwidth]{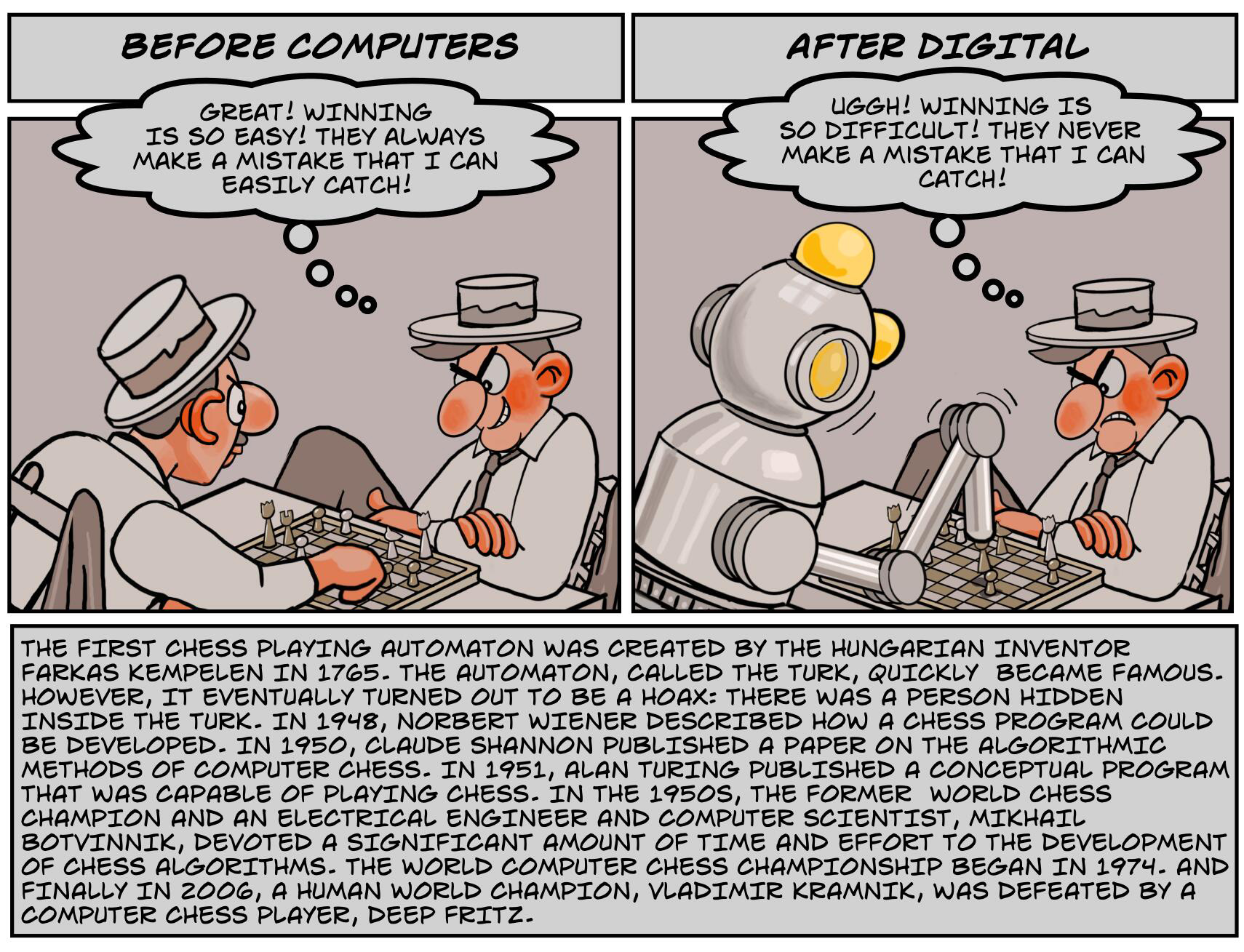}
        \caption{A human-generated cartoon published in the September 2024 issue of the IEEE Computer magazine \citep{akleman2024computing47}.}
        \label{fig_47o}
    \end{subfigure}
    \hfill
    \begin{subfigure}[b]{0.495\textwidth}
        \includegraphics[width=\textwidth]{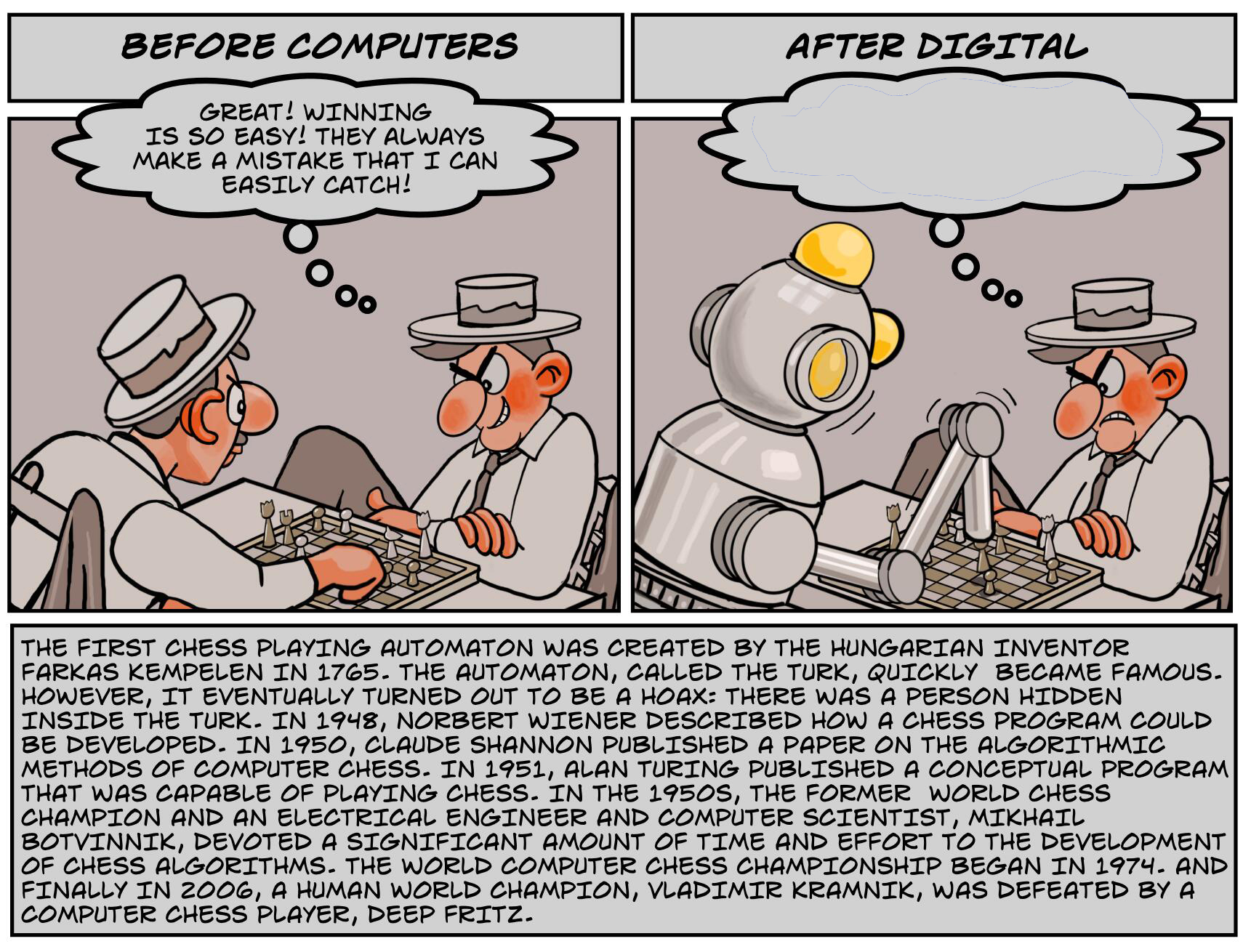}
        \caption{The same cartoon, but the speech balloon in the last panel is cleared so GPT-4o can fill it.}
        \label{fig_47}
    \end{subfigure}

    \caption{This study, inspired by the New Yorker caption contest \citep{allen2025new}, tests AI-generated cartoon captions. We removed the text from the second panel (\ref{fig_47o}), creating an image (\ref{fig_47}). \textcolor{black}{The GPT-4o model, accessed via the ChatGPT interface,} was prompted with \textit{"Fill the empty balloon with an appropriate joke in the provided comic strip!"}. The response was impressive: \textbf{"it's not fair! This machine never gets tired or distracted!"}. We requested two more jokes and received equally strong alternatives: \textit{"How am I supposed to win when it never needs a coffee break?"} and \textit{"I think it's cheating... who programmed this thing?"}}
    \label{fig:teaser}
\end{figure}

\begin{abstract} 
Crafting cartoon captions requires an understanding of humor, context, and the relationship between image and text. Traditionally, illustrators and writers collaborate to strengthen visual storytelling and comedic timing. With advances in natural language generation, Large Language Models (LLMs) can assist in this process. This study examines AI’s role in caption generation by testing GPT-4o via the ChatGPT interface on IEEE Computer magazine cartoons. By removing captions and prompting AI to generate replacements, we assess its ability to produce jokes that match the depicted situation and narrative intent. Our findings show that while AI-generated captions are often humorous and contextually relevant, they sometimes diverge from the cartoon’s intended meaning, for example by missing irony, cultural references, or contextual constraints. However, AI can also produce alternatives that broaden creative exploration and occasionally improve upon the original humor. We argue that current AI systems are best used as an assistant rather than a replacement for human creativity. By integrating AI-generated suggestions, cartoonists can explore diverse humor styles, streamline ideation, and refine final captions while retaining creative control. This study highlights AI’s potential as a practical tool for caption ideation within a hybrid human--AI workflow.
\end{abstract}

\twocolumn

\section{Introduction and Motivation}

\begin{figure}
\centering
  \includegraphics[width=0.495\textwidth]{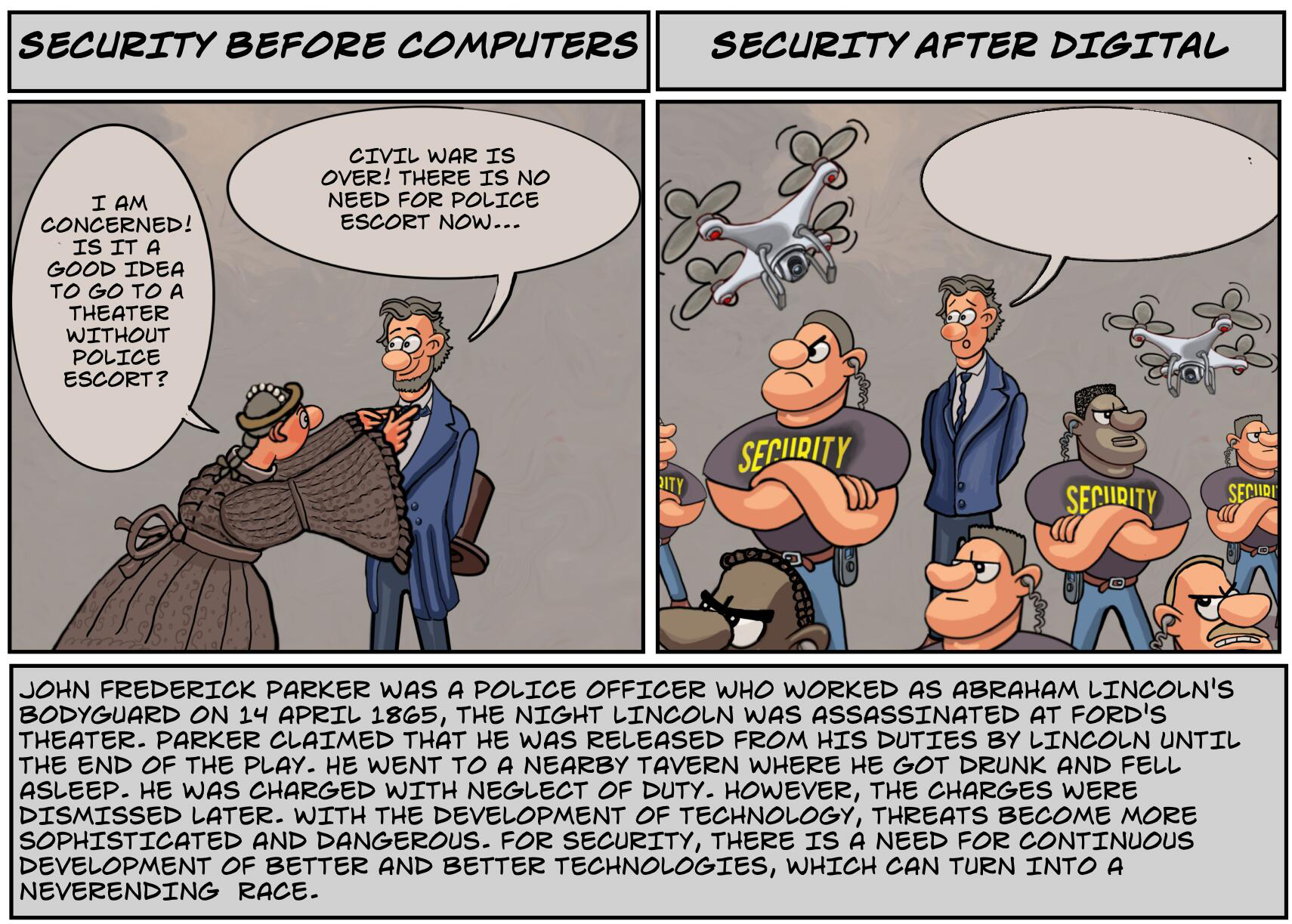}
  \caption{We provided GPT-4o this cartoon from IEEE Computer, published in October 2024 \citep{akleman2024computing48}, with the original text in the last speech balloon removed. GPT-4o generated the following three responses: \textit{(1) ``I think we might be overcompensating just a little!'' (2) ``So, does this come with a personal drone for my coffee runs?'' (3) ``I hope these drones know the difference between friends and enemies!''} The original caption was: \textit{``I am concerned! I definitely need more people and better technology for my security.''}}
\label{fig_48}
\end{figure}

\begin{figure}
\centering
  \includegraphics[width=0.495\textwidth]{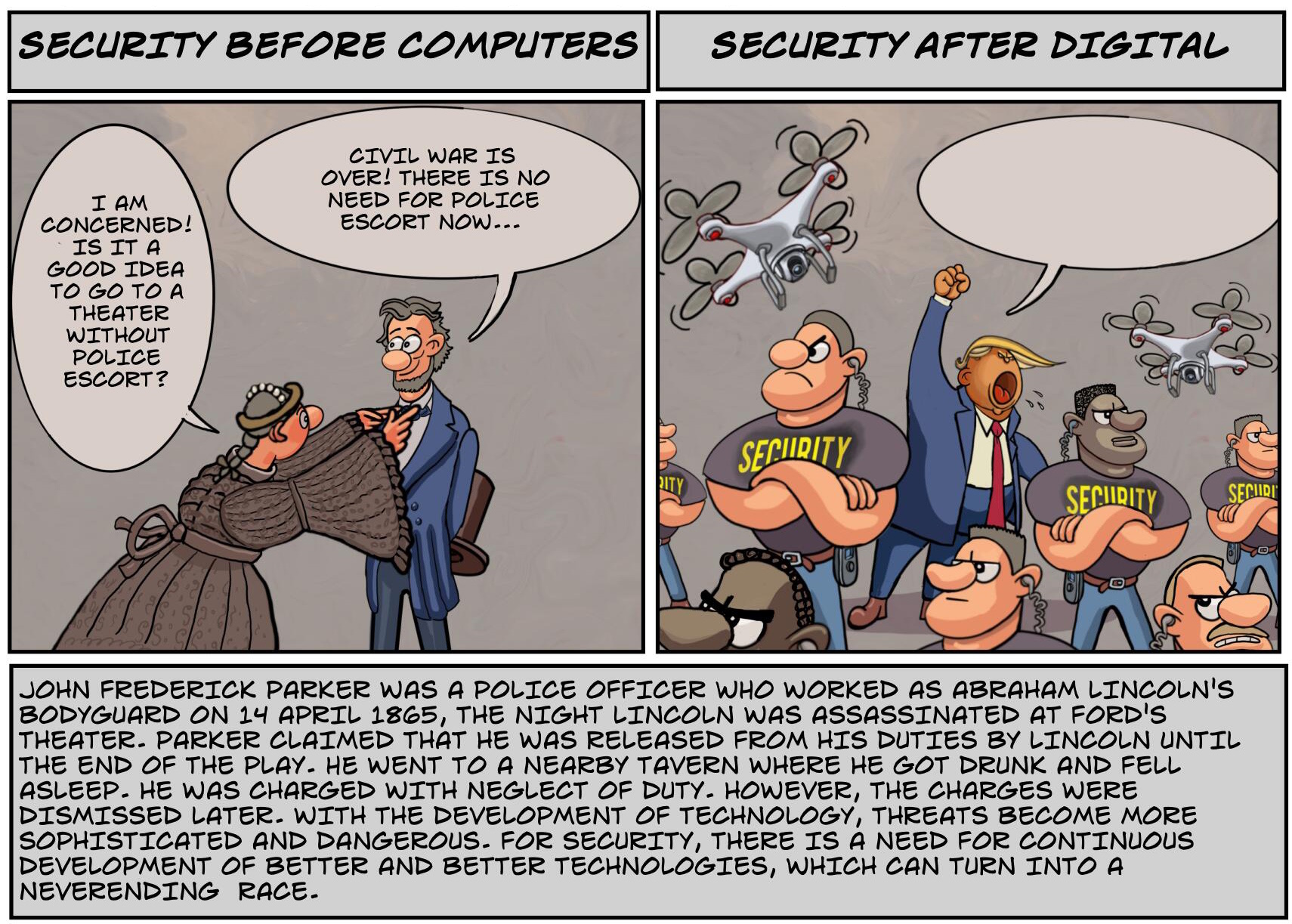}
  \caption{We provided GPT-4o the pre-publication version of the same cartoon shown in Figure~\ref{fig_48}, which originally featured a caricature of President Trump before being edited into a non-political version for publication in IEEE Computer \citep{akleman2024computing48}. Using the same procedure, we removed the original text in the last speech balloon. GPT-4o generated the following three responses: \textit{(1) ``We need to make America safe again... with more drones!'' (2) ``Make security great again!'' (3) ``I need a wall of drones, the best drones, nobody has better drones!''}. These responses indicate that GPT-4o successfully recognized and leveraged cultural references associated with the original caricature. The original caption was: \textit{``I am concerned! I definitely need more people and better technology for my security.''}}
\label{fig_48_trump}
\end{figure}

The use of artificial intelligence in content creation, and cartoon generation in particular, has emerged as a major topic of discussion in recent years. While cartoons may appear simple on the surface, producing visually engaging illustrations paired with effective humor demands substantial skill, experience, and creative judgment. Traditionally, cartoon production has followed two main models: either a single creator is responsible for both the drawing and the caption, or the task is divided between a writer and an illustrator working collaboratively. As AI systems capable of generating images and text have rapidly gained visibility, they have begun to challenge both of these long-standing creative paradigms.

Much of the current controversy surrounding AI in cartooning focuses on illustration. AI-based drawing and painting systems have intensified debates about authorship, originality, and artistic labor, with empirical work showing that audiences consistently perceive human-created art as having greater emotional depth, intentionality, and authenticity than AI-generated work \citep{vanHees2025,cunningham2025}. These concerns were amplified by recent high-profile incidents in which secretly AI-generated cartoons were submitted as original works and even won major competitions, triggering backlash once their origins were revealed \citep{cumhuriyet2024turhan,darroch2017netherland,kotbas2024yapay}. Critics often argue that such tools allow writers with little artistic training to bypass illustrators altogether.

However, collaboration between writers and illustrators is not a novel or controversial practice in itself. Many iconic comics and cartoons were created through precisely this division of labor, including Asterix by René Goscinny and Albert Uderzo \citep{goscinny1959asterix}, Lucky Luke by René Goscinny and Morris \citep{goscinny1978lucky}, and Hombre by Segura and Ortiz \citep{segura1981hombre}. The central question, then, is not whether collaboration is legitimate, but whether AI can meaningfully participate in such creative workflows. Prior investigations suggest that while AI-assisted illustration tools can produce visually plausible outputs, they lack the nuance, predictability, and intentional control characteristic of human artists, often requiring extensive trial-and-error to achieve acceptable results \citep{onal2025comparingperformance}.

\textcolor{black}{While collaboration between writers and illustrators is not inherently controversial, the introduction of AI into this relationship introduces distinct ethical and legal considerations. Unlike human collaborators, AI systems are not intentional agents and are often characterized as derivative systems trained on large-scale datasets \cite{Franceschelli2025, Runco02012025}, which raises ongoing debates about authorship, ownership, and creative responsibility \cite{Franceschelli2025, vanHees2025, moffat2023perception}. We acknowledge the importance of these discussions and recognize that human--AI collaboration differs fundamentally from human--human collaboration in this regard \cite{Wan2024, dellermann2019hybrid}. However, addressing these ethical and legal questions is not the primary objective of this study. Instead, our focus is on evaluating the practical capabilities of AI within a creative workflow, specifically its effectiveness as a tool for generating cartoon captions and supporting human creators in the ideation process.}

In this paper, we evaluate the less explored form of AI involvement in cartoon creation that aligns more closely with the traditional writer–illustrator model. Inspired by The New Yorker caption contest \citep{allen2025new}, we focus on the use of AI for caption generation rather than image creation. A strong caption is essential to a cartoon’s success, yet generating humorous, original, and well-matched text remains challenging even for experienced cartoonists. This difficulty is also reflected in recent large-scale studies showing that current AI systems still underperform human contributors in cartoon captioning tasks~\citep{zhang2024humor}. We investigate whether AI can function as a productive assistant in this second mode of cartoon creation by generating multiple caption candidates that human creators can evaluate, refine, and select from. By shifting the focus from replacing illustrators to supporting the writing process, \textcolor{black}{our study offers a more targeted evaluation of how AI can be integrated into established creative practices in cartooning, specifically by examining its capability to generate contextually appropriate and effective captions as part of a human-centered workflow.}

To evaluate AI’s capabilities, we removed captions from speech balloons in a series of cartoons and prompted GPT-4o via the ChatGPT interface to generate new text. \textcolor{black}{The AI-generated responses were frequently found to be humorous, contextually relevant, and well-aligned with the visual narrative by human evaluators in our user study.} These findings demonstrate that AI can function as an effective caption-generation tool, assisting cartoonists in finding the best possible text to enhance their illustrations.

By leveraging AI for caption suggestions, cartoonists can explore a broader range of humor styles, improve joke quality, and streamline their creative process. Rather than replacing human creativity, AI serves as a collaborative tool that strengthens the connection between visual storytelling and textual humor. Our findings underscore the potential of this hybrid model, where cartoonists retain artistic control while AI enhances their workflow by generating engaging captions.

\begin{figure}
\centering
  \includegraphics[width=0.495\textwidth]{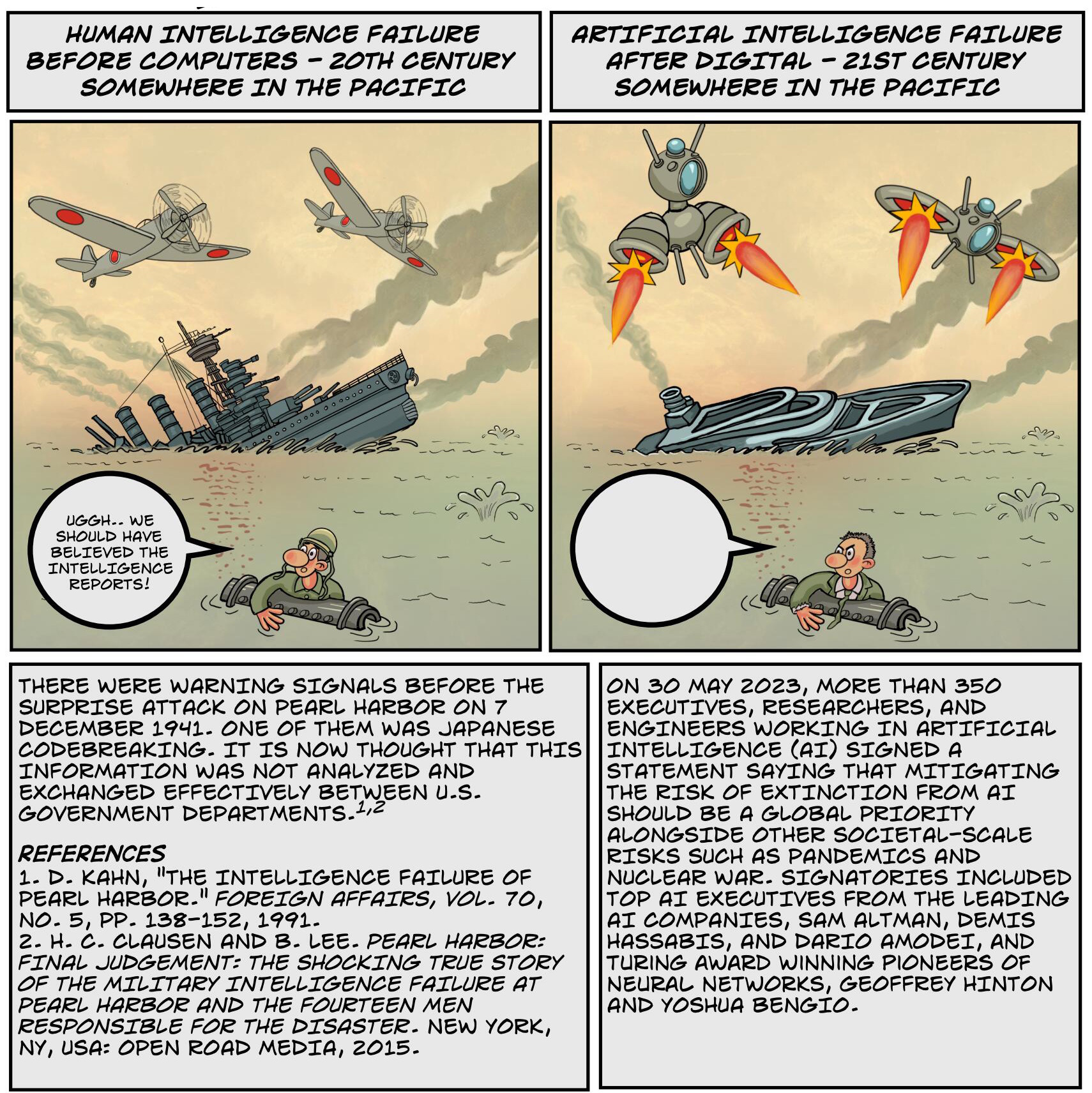}
        \caption{We provided GPT-4o this cartoon from IEEE Computer, published in November 2024 \citep{akleman2024computing49}, with the original text in the last speech balloon removed. GPT-4o generated the following three responses: \textit{(1) "Ughh... we should have updated the software!" (2) "I guess the AI didn’t get the memo about friendly fire!" (3) "Well, at least the AI is consistent... consistently wrong!"} While these jokes are amusing, they do not fully capture the concept of intelligence failure and double meaning of intelligence. The original caption was: \textit{"Ughh... we should have believed the AI expert warnings!}"}
        \label{fig_49}
\end{figure}

\begin{figure}
\centering
  \includegraphics[width=0.495\textwidth]{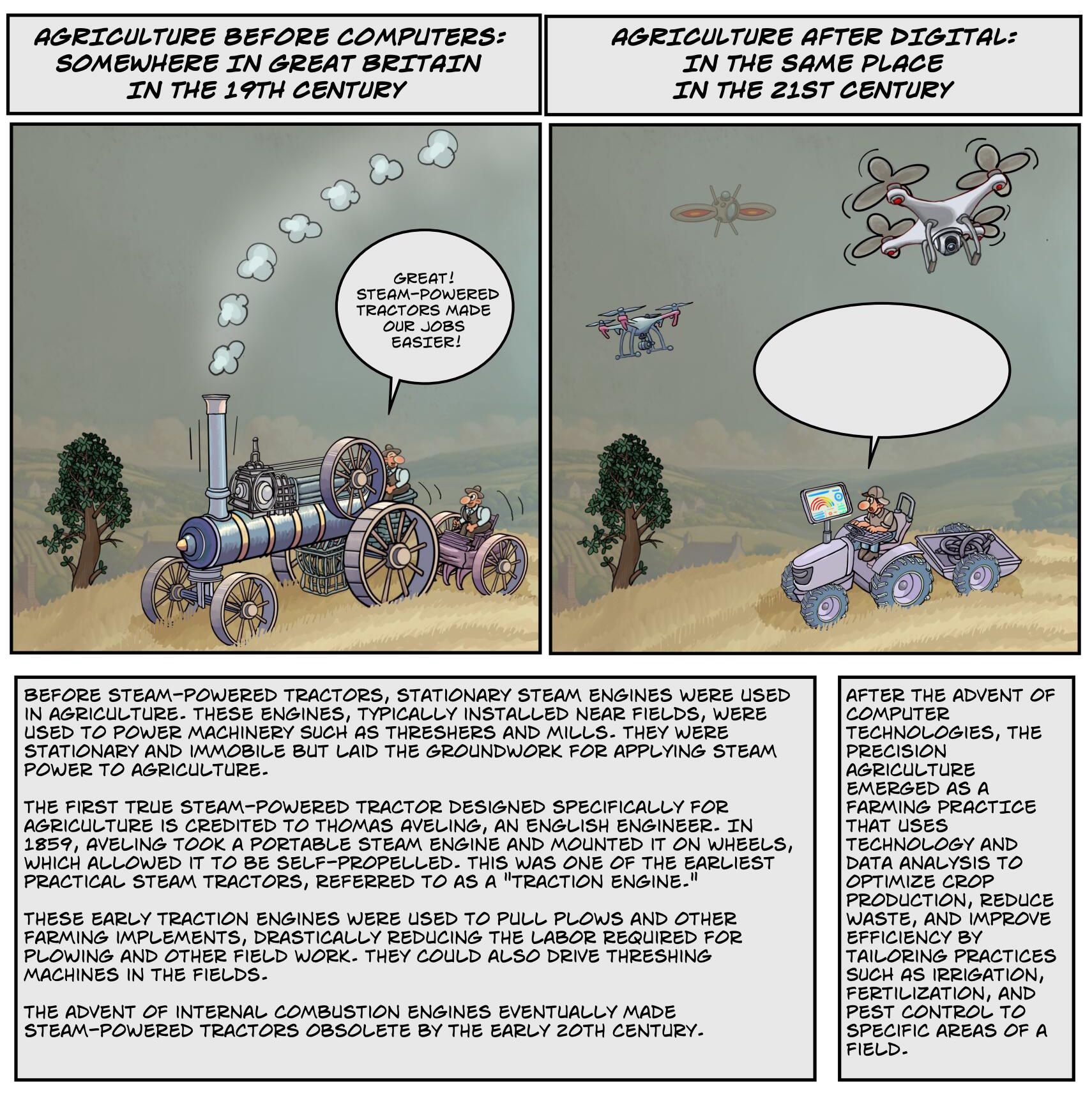}
        \caption{We provided GPT-4o this cartoon from IEEE Computer, published in December 2024 \citep{akleman2024computing50}, with the original text in the last speech balloon removed. GPT-4o generated the following three responses: \textit{(1) "These drones even tell me when to take a coffee break!" (2) "Finally, a farm that listens to me more than my kids do!" (3) "Now the drones work harder than we do!"} The original caption was: \textit{"Ughh... we should have also tracked down insect movements for better control of the damage they cause!"}
        }
        \label{fig_50}
\end{figure}

\subsection{Context and Motivation}
\label{Sec_ContextandMotivation}

Effective visual storytelling presents challenges even for seasoned professionals because meaning arises from how characters, actions, and setting combine into a coherent scene. Elements such as body posture or gaze direction can influence how emotions are perceived \citep{liu2012, akleman2015, dede2024power}, shaping how viewers interpret the situation. While AI is often heralded as a tool for enhancing efficiency and generating jobs requiring advanced skills \citep{trivedi2023should}, studies suggest that audiences favor human-created art for its emotional depth and perceived intentionality \citep{bellaiche2023humans, moffat2023perception}. Recent findings indicate that this preference remains robust even as generative models improve, with human-authored works continuing to be perceived as more intentional and emotionally grounded.

Despite its advancements, AI continues to exhibit notable shortcomings in graphic design and expressive imagery \citep{sindhura2021virtues}. More recent evaluations similarly report limitations in expressive control and semantic coherence in generative visual systems \citep{elgammal2023aesthetic, huang2024expressive}. A promising solution lies in hybrid intelligence, which leverages both human expertise and AI efficiency to optimize creative outputs. This approach has shown promise in domains such as celestial body classification via crowdsourcing \citep{kamar2012combining} and integrating expert knowledge with AI for improved outcomes \citep{chang2017revolt, bansal2021does}.

\textcolor{black}{The question of whether LLMs are genuinely creative has attracted considerable scholarly attention, spanning both empirical experiments and philosophical inquiry. Franceschelli and Musolesi~\cite{Franceschelli2025} argue that LLMs can at most exhibit \textit{combinatorial} or \textit{exploratory} creativity, recombining and extending patterns learned from large corpora, but fall short of \textit{transformational} creativity, which requires fundamentally restructuring the conceptual space itself. Empirical studies in domains such as poetry, story writing, and divergent association tasks show that LLMs approximate average human performance on psychometric measures such as the Divergent Association Task and Alternative Uses Test~\cite{BellemarePepin2026}, yet their outputs remain statistically more homogeneous than those of humans~\cite{Wenger2026}. From a philosophical standpoint, Runco and Jaeger~\cite{Runco02012025} emphasize that genuine creativity requires both \textit{authenticity} and \textit{intentionality}, a dual criterion that LLMs satisfy only partially, as their outputs depend on the intent and evaluative judgment of a human author. In the context of this work, AI is therefore best understood not as an autonomous creative agent, but as an iterative writing partner~\cite{Wan2024} that amplifies human creativity by rapidly generating diverse caption candidates, while the cartoonist retains creative agency over selection, intent, and authorship.}

In the context of cartoon captioning, our observations suggest that AI is most effective when integrated into the creative process as a support mechanism rather than as an autonomous author. Instead of producing a single definitive caption, AI-generated outputs offer a spectrum of alternatives that cartoonists can evaluate, refine, or selectively adapt to better align with visual intent and narrative tone. This multiplicity enables broader creative exploration, reduces iteration time, and exposes alternative humorous interpretations that may not immediately arise through human ideation alone. In this role, AI functions less as a creative decision-maker and more as a catalyst for exploration within a human-directed workflow.

\begin{figure}
\centering
  \includegraphics[width=0.495\textwidth]{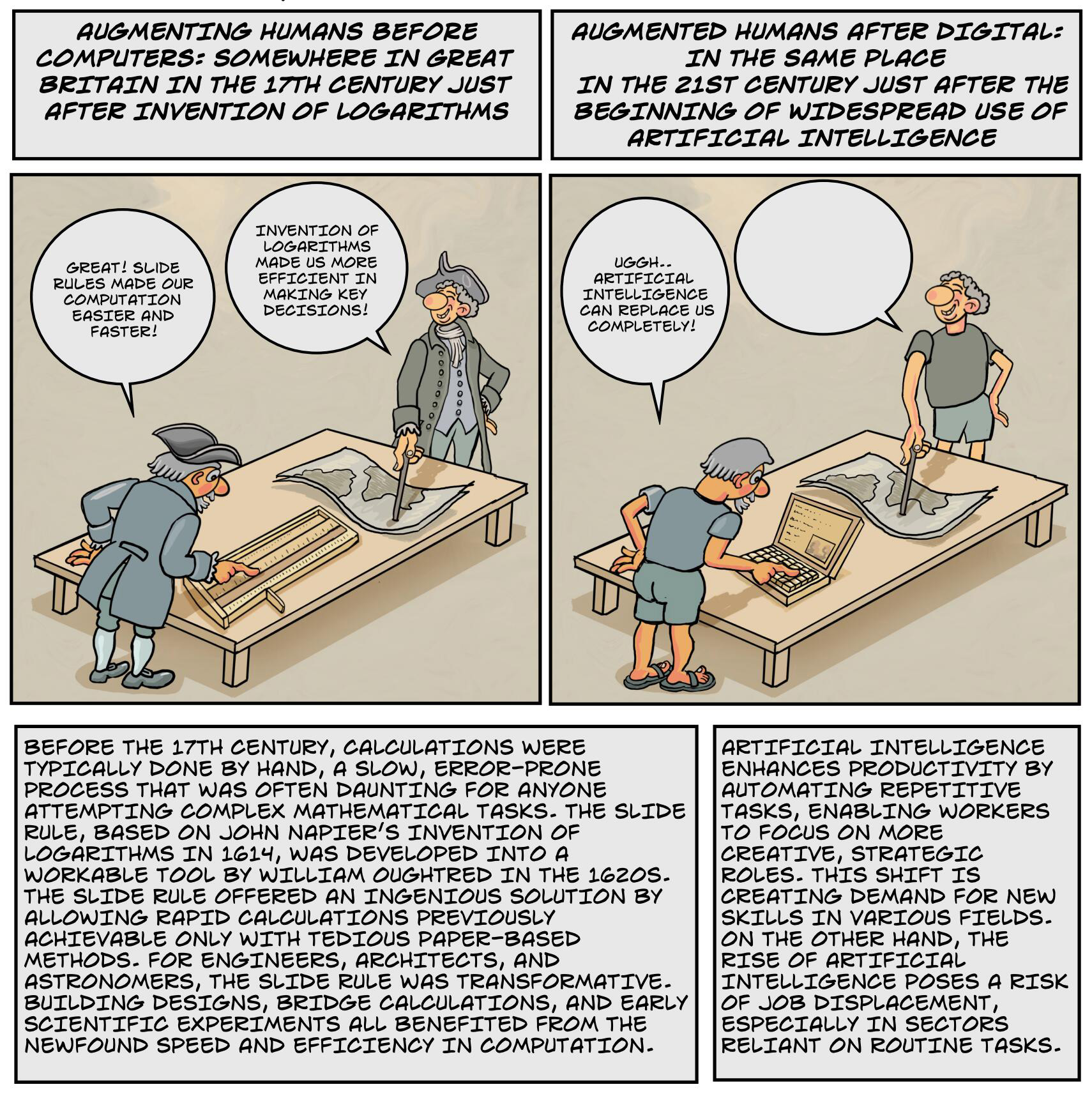}
        \caption{We presented GPT-4o with (\ref{fig_51}), a cartoon from IEEE Computer published in January 2025 \citep{akleman2025computing51}, with the original text in the last speech balloon removed. GPT-4o generated the following three responses: \textit{(1) "At least we don't have to worry about slide rule maintenance anymore!" (2)  "Well, at least AI doesn’t complain about work hours!" (3) "I guess it's time to teach AI how to take coffee breaks!"} These are lighthearted and amusing responses, but they do not fully capture the main concept. The original caption was:\textit{ "No! It made us more efficient in making key decisions!"}
        }
        \label{fig_51}
\end{figure}

\begin{figure}
\centering
  \includegraphics[width=0.495\textwidth]{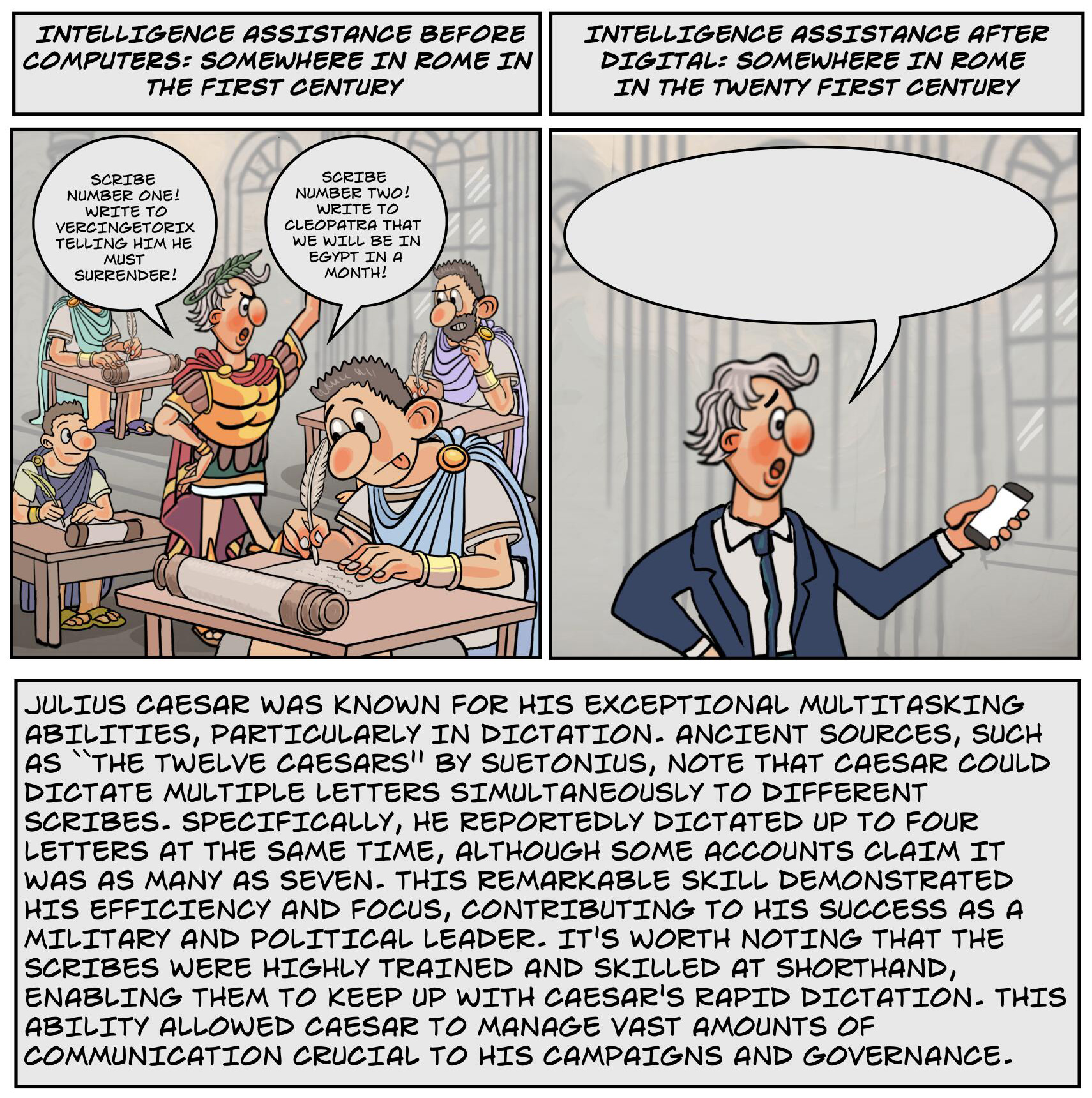}
        \caption{We presented GPT-4o with (\ref{fig_52}), a cartoon from IEEE Computer published in February 2025 \citep{akleman2025computing52}, with the original text in the last speech balloon removed. GPT-4o generated the following three responses: \textit{(1) "Siri, send a letter to the Senate… and order me some pizza!" (2) "Alexa, conquer Gaul and set a reminder for lunch!" (3) "ChatGPT, write my speech… and make it sound historic!"}. These responses are lighthearted and amusing but do not fully capture the main concept. The original caption was: \textit{"Write 'Let's meet at this great restaurant that just opened nearby.' Now send it to... Ughh, I forgot whether I sent this to my co-workers through WhatsApp, my high school friends through Facebook, or my college roommates via email!"}}
        \label{fig_52}
\end{figure}

\subsection{Basis, Rationale, and Contributions}
\label{Sec_BasisAndRationale}

The scientific community has yet to fully elucidate the mechanisms by which people interpret expressive imagery and narrative intent. However, expert actors, comic book artists, cartoonists, and animators have long developed practical methods for constructing readable scenes and guiding audience interpretation. This expertise is typically acquired through apprenticeship and mentorship rather than formal documentation. Only a handful of artists have explicitly described their methodologies \citep{johnston1981illusion, mccloud1993understanding, blair1995cartoon, mccloud2006making, eisner2008comics, eisner2008graphic, celik2011on}. Conventional user study procedures can also be difficult to apply when the goal is to isolate why a caption works or fails in a particular scene. This motivates a simplified evaluation setting that allows rapid iteration and comparison.

Caption generation provides such a setting. Because the illustration remains fixed and human-authored, the creative decision space is constrained to the textual layer. This makes it possible to test whether an AI system can (i) understand what is happening in the scene at a high level and (ii) propose multiple caption candidates that a cartoonist can evaluate, refine, or reject.

Our study showed that GPT-4o often produces humorous captions that are consistent with the depicted situation. At the same time, it can diverge from the cartoon’s intended meaning, especially when the humor depends on irony, cultural references, or implicit context. \textcolor{black}{These patterns reinforce that AI is best positioned as an assistant that expands the set of candidate captions, while human creators retain authority over final selection and authorial intent. This finding aligns with philosophical accounts of LLM creativity, which argue that LLMs can achieve combinational and exploratory creativity but not transformational creativity \citep{Franceschelli2025}, and that LLM outputs, while potentially original and effective, lack the authenticity and intentionality characteristic of human-authored work \citep{Runco02012025}.}

%Additionally, ethical considerations must be accounted for when integrating AI into creative domains. AI-generated humor has the potential to introduce biases, misrepresent cultural nuances, or unintentionally reinforce stereotypes. Ensuring fairness and diversity in AI training data is crucial to preventing problematic outputs. Furthermore, the question of authorship remains an important debate—while AI can generate captions, the creative agency and final artistic decisions should remain with human cartoonists to preserve originality and artistic intent.

Crucially, for our hypothesis to hold, it is sufficient to demonstrate that at least one AI system can support this role. In this study, we tested this hypothesis using GPT-4o, and our experimental findings affirmed its validity. The AI-generated captions frequently aligned with the visual narrative and, in some cases, improved clarity or impact. However, human intervention remained necessary to refine and select the best options. This supports AI-assisted storytelling as a hybrid workflow in which AI broadens ideation and the cartoonist preserves creative agency.

\begin{figure}
\centering
  \includegraphics[width=0.495\textwidth]{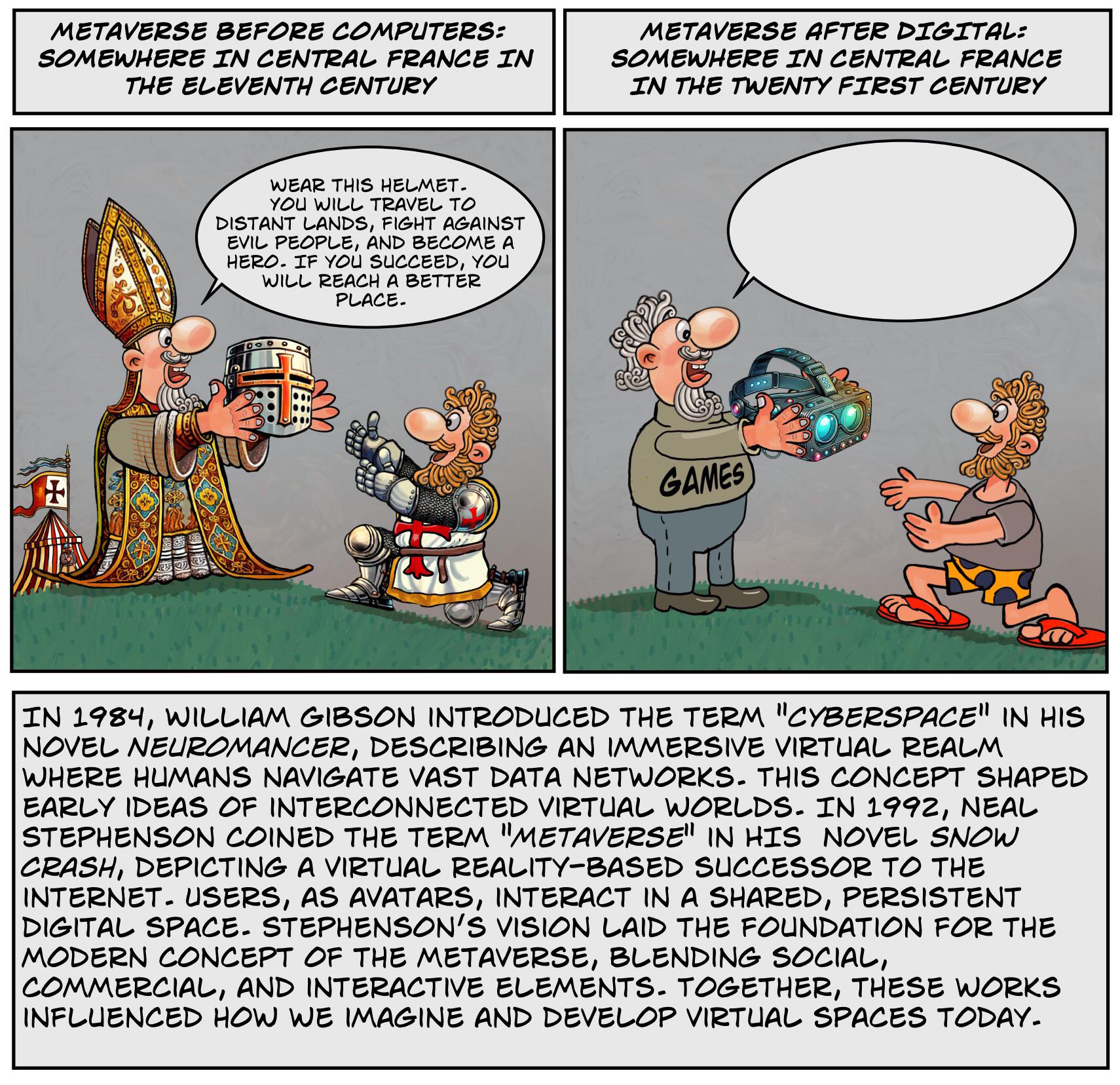}
        \caption{We presented GPT-4o with (\ref{fig_53}), a cartoon from IEEE Computer published in March 2025 \citep{akleman2025computing53}, with the original text in the last speech balloon removed. GPT-4o generated the following three responses: \textit{(1) "Put this on, and you can fight dragons without leaving your couch!" (2) "With this, you can conquer worlds… and still be home for dinner!" (3) "It’s like a quest, but with better graphics and no real danger!"} These responses somewhat captured the main concept. The original caption was: \textit{"Wear this helmet. You will travel to faraway lands, fight against bad people, and become a hero. If you are successful, you will advance to the next level."}}
        \label{fig_53}
\end{figure}

\section{Methodology}
\label{sec:methodology}

Our evaluation follows a four-step pipeline: (1) selecting a structured family of cartoons, (2) removing the text from the final speech balloon in each cartoon, (3) prompting an AI system to generate multiple candidate captions \textcolor{black}{(using a consistent prompt, as described in Figure~\ref{fig:teaser} and Section~\ref{subsec:data})}, and (4) evaluating these captions against the original ones through a survey-based user study.

\subsection{Experiment Design}
\label{subsec:experiment-design}

We employed a survey-based experimental design to examine how audiences perceive AI-generated cartoon captions in comparison to human-authored ones. Participants were presented with cartoons in which the final speech balloon contained either the original caption written by the cartoonist or an AI-generated alternative. \textcolor{black}{For each cartoon, participants evaluated exactly two captions: one human-authored and one randomly selected AI-generated caption.} The study was designed to measure both subjective preference and participants’ ability to identify AI authorship.

\paragraph{Survey Questions.}
For each cartoon, participants answered the following questions:

\begin{itemize}
\item \textit{Which cartoon's last speech bubble do you think contains text generated by AI?}
\item \textit{Which joke do you prefer?}
\item \textit{What would you write inside the last speech bubble?} (Optional)
\end{itemize}

The first question captures participants’ perception of AI authorship, while the second measures caption preference. The optional third question was included to encourage deeper engagement with the cartoon and to observe how participants might reinterpret or extend the presented humor.

\paragraph{Survey Variants and Question Order.}
To account for potential order effects and priming related to AI authorship, we implemented three survey variants that differed only in the ordering and separation of these questions:

\begin{itemize}
\item \textcolor{black}{\textbf{Guess First (GF):} Participants first guessed which caption was generated by AI and then indicated which caption they preferred. These two questions were presented consecutively for each cartoon before moving on to the next.}
\item \textcolor{black}{\textbf{Preference First (PF):} Participants first selected their preferred caption and then guessed which one was generated by AI. These two questions were presented consecutively for each cartoon before moving on to the next.}
\item \textcolor{black}{\textbf{Separated (SEP):} Participants first viewed all cartoons and indicated their preferred captions in a single continuous section. In a separate section, they then viewed the same cartoons again to guess AI authorship. Demographic questions were presented at the end.}
\end{itemize}

These variants were designed to assess whether explicitly considering AI authorship influences preference judgments. In particular, the Separated variant was intended to minimize priming effects by fully decoupling preference evaluation from authorship attribution. The use of multiple survey variants was motivated by prior findings in survey methodology and cognitive research showing that question order and task sequencing can systematically influence subjective judgments. Experimental studies demonstrate that earlier questions can prime respondents’ attention and bias subsequent evaluations, even when the underlying stimuli remain unchanged \citep{Papp2024, Novella2024}. More recent work further indicates that survey flow affects self-reported cognitive and evaluative measures, highlighting the role of contextual framing in shaping responses \citep{Donmez2025}. These findings motivated our inclusion of the SEP variant, which decouples preference evaluation from authorship attribution to reduce origin-focused priming.

\subsection{Data}
\label{subsec:data}

The primary challenge with this approach is gaining access to multiple families of cartoons due to copyright restrictions. To address this, we selected a structured family of cartoons published in the flagship publication of the IEEE Computer Society, IEEE Computer magazine \citep{akleman2024computing50,akleman2024computing49,akleman2024computing47,akleman2024computing48,akleman2025computing51,akleman2025computing52,akleman2025computing53}. Using these cartoons offers several advantages.

First, the dataset is substantial, with more than 50 published cartoons available. Second, all these cartoons are accessible through IEEE Computer magazine and can be officially cited, as they have unique digital object identifiers within IEEE’s archives. Third, the jokes in these cartoons are directly inspired by the magazine’s main topic for that particular month, ensuring a strong connection to contemporary technological issues.

Most importantly, the fourth and most significant advantage is that these cartoons follow a consistent structural format across every issue, making them ideal for systematic evaluation. 
This structure established in 2018 by IEEE Computer Editor Sumi Helal and cartoonist Ergun Akleman is still used in all cartoons today. These cartoons consist of two nearly identical panels, titled "Before Computers" and "After Digital."

In the first panel, characters are typically delighted by a technological advancement and express their excitement with a positive statement that generally begins with "Great!" In the second panel, despite further advancements with computers, the characters always desire more and express dissatisfaction with a negative remark generally starting with "Ughh!" This consistent structure highlights the evolving relationship between technology and human expectations.

For this study, we selected seven IEEE Computer cartoons: \textit{Playing Chess} (Figure~\ref{fig_47}), \textit{Security} (Figure~\ref{fig_48}), \textit{Intelligence Failures} (Figure~\ref{fig_49}), \textit{Agriculture} (Figure~\ref{fig_50}), \textit{Augmented AI} (Figure~\ref{fig_51}), \textit{Intelligence Assistance} (Figure~\ref{fig_52}), and \textit{Metaverse} (Figure~\ref{fig_53}). In addition, we included a pre-publication variant of one cartoon that was not published due to its political content. For each cartoon, we removed the text from the final speech balloon and prompted GPT-4o to generate three alternative captions using the same sequence of prompts for every image. \textcolor{black}{The same instruction prompt was used consistently across all cartoons: \textit{"Fill the empty balloon with an appropriate joke in the provided comic strip!"}.} All three responses are included in the figure captions. As shown in the examples, all GPT-4o-generated responses are useful, with some being exceptionally well-crafted.

In particular, we want to highlight the responses generated for Figure~\ref{fig_48_trump}. GPT-4o correctly identified that the speaker in the second panel was President Donald Trump and crafted jokes by creatively manipulating his signature phrases. One example is:
\textit{"We need to make America safe again... with more drones!"}
Notably, the ellipsis ("...") creates a long pause before the unexpected twist, "with more drones," enhancing the comedic effect. Another example follows a similar pattern:
\textit{"Make security great again!"}
Finally, the last response directly references another well-known Trump phrase:
\textit{"I need a wall of drones, the best drones—nobody has better drones!"}
These examples illustrate GPT-4o’s ability to recognize context and generate humor by adapting well-known speech patterns.

\textcolor{black}{All caption generation was performed using GPT-4o via the ChatGPT interface during our data collection period, which concluded on February 25, 2025. At that time, several of the cartoons used in this study (including those shown in Figures~\ref{fig_52} and \ref{fig_53}) had not yet been published, preventing any possibility of retrieving their original captions from external sources. Additionally, one cartoon corresponds to a pre-publication version that was never released in that form (see Figure~\ref{fig_48_trump}). GPT-4o’s training data is understood to extend up to approximately late 2023; therefore, these cartoons would not have been present in the model’s training corpus. While ChatGPT may support web-browsing capabilities when explicitly invoked, this functionality was not used in our experiments. All prompts were issued in a controlled setting without enabling external search, ensuring that the outputs are derived solely from the model’s internal representations.}

\subsection{Participant Demographics}
\label{subsec:survey-demographics}

Table~\ref{tab:demo-age} reports the age distribution of participants. The sample was predominantly composed of young adults, with the majority of participants in the 25--34 age group, followed by those aged 18--24. Older age groups were comparatively less represented, which is consistent with the recruitment context and the target population of digitally engaged participants. Throughout the demographic analysis, \(N\) denotes the total number of participants, while \(n\) denotes the number of participants within each category.

\begin{table}[ht]
  \centering
  \caption{Age group distribution (participant-level, \(N=66\)). Values are \(n~(\%)\).}
  \label{tab:demo-age}
  \scriptsize
  \begin{tabular}{lc}
    \toprule
    Age Group & Count (\%) \\
    \midrule
    18--24    & \textbf{26} (39.4) \\
    25--34    & \textbf{34} (51.5) \\
    35--44    & \textbf{6}  (9.1) \\
    \bottomrule
  \end{tabular}
\end{table}

Interest and expertise were captured across three topics: Artificial Intelligence (AI), Arts and Illustrations, and Comics and Cartoons. Tables~\ref{tab:demo-interest-all} and \ref{tab:demo-expertise-all} summarize the distributions of self-reported interest levels and expertise levels, respectively. Overall, participants reported higher interest and expertise in AI compared to the two art-related domains, while Arts and Illustrations and Comics and Cartoons exhibited broader distributions skewed toward lower to intermediate expertise levels. This diversity indicates a heterogeneous participant pool with varying backgrounds across technical and creative domains. \textcolor{black}{Our study focuses on perception-based measures, specifically caption preference and perceived authorship, which are inherently audience-dependent rather than dependent on domain expertise. In this context, a broad, digitally engaged audience provides meaningful insight into how AI-generated captions are received in practice. Expert evaluation could provide additional insights into nuanced aspects of humor, authorship, and stylistic quality; however, such an evaluation targets different research questions than the ones addressed here.}

\begin{table}[ht]
  \centering
  \caption{Interest by topic (participant-level, \(N=66\)). Values are \(n~(\%)\).}
  \label{tab:demo-interest-all}
  \scriptsize
  \resizebox{0.495\textwidth}{!}{%
      \begin{tabular}{lccc}
        \toprule
        Level & AI & Arts and Illustrations & Comics and Cartoons \\
        \midrule
        None                         & \textbf{4}  (6.1)  & \textbf{4}  (6.1)  & \textbf{7}  (10.6) \\
        Somewhat Interested          & \textbf{9}  (13.6) & \textbf{18} (27.3) & \textbf{18} (27.3) \\
        Interested                   & \textbf{21} (31.8) & \textbf{22} (33.3) & \textbf{18} (27.3) \\
        Very Interested              & \textbf{21} (31.8) & \textbf{12} (18.2) & \textbf{15} (22.7) \\
        Passionate / Deeply Interested & \textbf{11} (16.7) & \textbf{10} (15.2) & \textbf{8}  (12.1) \\
        \bottomrule
      \end{tabular}
  }
\end{table}

\begin{table}[ht]
  \centering
  \caption{Expertise by topic (participant-level, \(N=66\)). Values are \(n~(\%)\).}
  \label{tab:demo-expertise-all}
  \resizebox{0.495\textwidth}{!}{%
      \begin{tabular}{lccc}
        \toprule
        Level & AI & Arts and Illustrations & Comics and Cartoons \\
        \midrule
        No Experience & \textbf{6}  (9.1)  & \textbf{22} (33.3) & \textbf{25} (37.9) \\
        Beginner      & \textbf{19} (28.8) & \textbf{18} (27.3) & \textbf{21} (31.8) \\
        Intermediate  & \textbf{23} (34.8) & \textbf{12} (18.2) & \textbf{12} (18.2) \\
        Advanced      & \textbf{11} (16.7) & \textbf{13} (19.7) & \textbf{6}  (9.1) \\
        Expert        & \textbf{7}  (10.6) & \textbf{1}  (1.5)  & \textbf{2}  (3.0) \\
        \bottomrule
      \end{tabular}
  }
\end{table}

\section{Survey Results}
\label{sec:survey-results}

As shown in Table~\ref{tab:overall-and-variant}, across all \(462\) decisions (66 participants \(\times\) 7 cartoons), AI captions were preferred in 300 cases (64.9\%), meaning that in nearly two out of three comparisons, the participants favored the AI's output over the human-written original caption. Participants correctly identified the AI involvement in 141 decisions (30.5\%), indicating that, in most cases, the origin of the caption was not obvious. This combination of high AI preference and low identification accuracy suggests that many AI-generated captions were evaluated positively on their own merits rather than being influenced by perceptions of machine authorship. In the survey results, \(N\) refers to the total number of caption-comparison decisions, while \(n\) indicates the number of decisions per outcome or per cartoon.

\textcolor{black}{To examine whether these aggregate patterns differ across survey variants, we also analyzed results separately for the GF, PF, and SEP conditions, as also summarized in Table~\ref{tab:overall-and-variant}. AI captions were preferred at comparable rates across variants: GF (64.93\%), PF (62.86\%), and SEP (63.91\%). Correct AI identification similarly remained low across variants: GF (30.52\%), PF (31.43\%), and SEP (22.56\%). These results confirm that the aggregated findings are not driven by any single survey design, and instead reflect a robust pattern across different task orderings. Variant-specific effects on the interplay between preference and perceived authorship are examined further through the regression analyses below.}

\begin{table}[ht]
  \centering
  \caption{\textcolor{black}{Decision-level AI caption preference and correct identification rates, overall and by survey variant. Values are \(n~(\%)\).}}
  \label{tab:overall-and-variant}
  \scriptsize
  \resizebox{0.495\textwidth}{!}{%
  \begin{tabular}{lcc}
    \toprule
    Condition & AI Caption Preference & Correct AI Identification \\
    \midrule
    Guess First (GF)      & \textbf{91}  (64.93\%) & \textbf{43}  (30.52\%) \\
    Preference First (PF) & \textbf{88}  (62.86\%) & \textbf{44}  (31.43\%) \\
    Separated (SEP)       & \textbf{121} (63.91\%) & \textbf{54}  (22.56\%) \\
    \midrule
    Total                 & \textbf{300} (64.94\%) & \textbf{141} (30.52\%) \\
    \bottomrule
  \end{tabular}
  }
\end{table}

Table~\ref{tab:per-cartoon} reports item-level results. Five of seven cartoons (\textit{Security}, \textit{Intelligence Failures}, \textit{Agriculture}, \textit{Augmented AI}, \textit{Intelligence Assistance}) show a clear majority preference for AI captions. \textit{Playing Chess} is mixed, and \textit{Metaverse} favors the human caption. Identification accuracy is low overall.

\begin{table}[h!]
  \centering
  \caption{Decision-level results by cartoon (66 decisions per cartoon). Values are \(n~(\%)\).}
  \label{tab:per-cartoon}
  \resizebox{0.495\textwidth}{!}{%
      \begin{tabular}{lccc}
        \toprule
        Cartoon & AI Caption Preference & Correct AI Identification \\
        \midrule
        Playing Chess            & \textbf{29} (43.9) & \textbf{30} (45.5) \\
        Security                 & \textbf{51} (77.3) & \textbf{15} (22.7) \\
        Intelligence Failures    & \textbf{44} (66.7) & \textbf{19} (28.8) \\
        Agriculture              & \textbf{47} (71.2) & \textbf{12} (18.2) \\
        Augmented AI             & \textbf{57} (86.4) & \textbf{8}  (12.1) \\
        Intelligence Assistance  & \textbf{48} (72.7) & \textbf{32} (48.5) \\
        Metaverse                & \textbf{24} (36.4) & \textbf{25} (37.9) \\
        \bottomrule
      \end{tabular}
  }
\end{table}

To formally examine the relationship between guessing and preference, we fitted a series of binary logistic regression models. In the first model, Preference was treated as the dependent variable with Variant (GF, PF, SEP), Guess, and their interaction as predictors (Table~\ref{tab:reg_pref_variant}). The model fit was significantly better than the null model ($p < .001$). A logistic regression analysis tested whether caption variant (GF, PF, SEP), participants’ authorship guess (AI vs. human), and their interaction predicted preference. GF and “human” guesses served as the reference categories. Importantly, authorship was not disclosed; analyses are based on participants’ perceptions of whether a caption was AI- or human-generated.

There was a strong main effect of perceived authorship. Captions that participants guessed were AI-generated were substantially less likely to be preferred than captions they guessed were human-written (OR = 0.117, 95\% CI [0.059, 0.232], $p < .001$), corresponding to an approximately 88\% reduction in the odds of preference.

Caption variant also influenced preference, conditional on perceived authorship. When captions were perceived as human-written, PF captions were significantly less likely to be preferred than GF captions (OR = 0.453, 95\% CI [0.237, 0.866], $p = .017$). SEP captions showed a similar but marginal pattern (OR = 0.525, 95\% CI [0.275, 1.003], $p = .051$).

Crucially, the effect of perceived authorship differed by variant. The PF × Guess interaction was significant (OR = 3.386, 95\% CI [1.237, 9.271], $p = .018$), indicating that the reduction in preference associated with guessing “AI” was attenuated for PF captions relative to GF captions. In contrast, the SEP × Guess interaction was not significant (OR = 1.599, 95\% CI [0.525, 4.869],$ p = .408$), suggesting that the perceived AI penalty for SEP captions did not reliably differ from that for GF captions.

Overall, participants were much less likely to prefer captions they believed were AI-generated, and this perceived authorship penalty varied by caption type for PF, but not SEP, captions.

\begin{table}[h]
\centering
\caption{Logistic regression results predicting Preference (DV) from caption Variant (GF, PF, SEP), Guess, and their interaction. Odds ratios (OR) and 95\% confidence intervals are reported for interpretability.}
\label{tab:reg_pref_variant}
\resizebox{0.495\textwidth}{!}{%
\begin{tabular}{lccccccc}
\toprule
Predictor & $\beta$ & SE & $z$ & $p$ & OR & 95\% CI (OR) \\
\midrule
Intercept (GF) & 1.629 & 0.245 & 6.662 & $<.001$ & 5.100 & [3.158, 8.236] \\
PF & $-0.792$ & 0.330 & $-2.396$ & .017 & 0.453 & [0.237, 0.866] \\
SEP & $-0.644$ & 0.330 & $-1.952$ & .051 & 0.525 & [0.275, 1.003] \\
Guess (AI) & $-2.148$ & 0.352 & $-6.110$ & $<.001$ & 0.117 & [0.059, 0.232] \\
PF $\times$ Guess & 1.220 & 0.514 & 2.373 & .018 & 3.386 & [1.237, 9.271] \\
SEP $\times$ Guess & 0.470 & 0.568 & 0.827 & .408 & 1.599 & [0.525, 4.869] \\
\bottomrule
\end{tabular}
}
\end{table}

A complementary model was then fitted with Guess as the dependent variable and Variant, Preference, and their interaction as predictors (Table~\ref{tab:reg_guess_variant}). This model also demonstrated good fit relative to the null ($p < .001$). A logistic regression analysis examined whether caption variant (GF, PF, SEP), preference, and their interaction predicted participants’ authorship guesses (AI vs. human). GF and non-preferred captions served as the reference categories. As in the previous Model, authorship was not disclosed; the dependent variable reflects participants’ perceptions of whether a caption was AI-generated.

There was a strong main effect of preference. Captions that participants preferred were substantially less likely to be guessed as AI-generated than captions they did not prefer (OR = 0.117, 95\% CI [0.059, 0.232], $p < .001$), corresponding to an approximately 88\% reduction in the odds of an AI guess.

Caption variant also influenced perceived authorship. Among non-preferred captions, PF captions were significantly less likely than GF captions to be judged as AI-generated (OR = 0.378, 95\% CI [0.176, 0.810], $p = .012$). SEP captions showed a similar pattern (OR = 0.340, 95\% CI [0.155, 0.744], $p = .007$), indicating lower odds of being labeled AI relative to GF captions.
The relationship between preference and perceived authorship varies by variant. The PF × Preference interaction was significant (OR = 3.386, 95\% CI [1.237, 9.271], $p = .018$), indicating that the negative association between preference and AI guesses was attenuated for PF captions compared to GF captions. In contrast, the SEP × Preference interaction was not significant (OR = 1.599, 95\% CI [0.525, 4.869], $p = .408$), suggesting that the link between preference and perceived authorship did not reliably differ between SEP and GF captions.

Overall, participants were much less likely to judge captions as AI-generated when they preferred them, and this association was moderated by caption type for PF, but not SEP, captions.

\begin{table}[h]
\centering
\caption{Logistic regression results predicting Guess (DV) from caption Variant (GF, PF, SEP), Preference, and their interaction. Odds ratios (OR) and 95\% confidence intervals are reported for interpretability.}
\label{tab:reg_guess_variant}
\resizebox{0.495\textwidth}{!}{%
\begin{tabular}{lccccccc}
\toprule
Predictor & $\beta$ & SE & $z$ & $p$ & OR & 95\% CI (OR) \\
\midrule
Intercept (GF) & 0.742 & 0.272 & 2.731 & .006 & 2.100 & [1.233, 3.577] \\
PF & $-0.974$ & 0.390 & $-2.499$ & .012 & 0.378 & [0.176, 0.810] \\
SEP & $-1.078$ & 0.399 & $-2.700$ & .007 & 0.340 & [0.155, 0.744] \\
Preference & $-2.148$ & 0.352 & $-6.110$ & $<.001$ & 0.117 & [0.059, 0.232] \\
PF $\times$ Preference & 1.220 & 0.514 & 2.373 & .018 & 3.386 & [1.237, 9.271] \\
SEP $\times$ Preference & 0.470 & 0.568 & 0.827 & .408 & 1.599 & [0.525, 4.869] \\
\bottomrule
\end{tabular}
}
\end{table}

Across both modeling directions, preference and perceived authorship were strongly associated: captions guessed to be AI-generated were substantially less likely to be preferred, and preferred captions were substantially less likely to be judged as AI-generated. The caption variant showed a limited moderation of this relationship. Specifically, the association between preference and AI guessing was attenuated for PF captions relative to GF captions, whereas no reliable moderation emerged for SEP captions. Together with the descriptive results, these findings suggest that evaluations and authorship perceptions were closely intertwined, with participants tending to infer human authorship captions they liked.

\textcolor{black}{Taken together with the overall and per-cartoon analyses, the order-effect findings indicate that the close relationship between preference and perceived authorship is robust across survey designs. Although participants tended to penalize captions they believed to be AI-generated, their ability to accurately identify authorship remained low, suggesting that AI and human captions were often difficult to distinguish. At the same time, AI-generated captions were frequently preferred and, in many cases, matched or exceeded human-written captions in audience evaluations. This combination of high preference and low identification accuracy indicates that the appeal of AI-generated captions often rests on perceived content quality rather than clearly distinguishable stylistic differences, supporting their potential as a creative collaborator in comic caption ideation.}

\section{Conclusion and Future Work}

Our study demonstrates that AI, particularly GPT-4o, can effectively generate captions for cartoons by providing contextually relevant and humorous responses. By systematically testing AI-generated captions against professionally crafted ones, we observed that AI is capable of producing compelling and amusing text that aligns with the narrative of the cartoons. While AI-generated captions do not always perfectly match the intended humor of the original captions, they frequently offer creative alternatives that enhance the overall storytelling process.

A key takeaway from our findings is that AI can serve as a valuable tool for cartoonists, helping them refine jokes and explore alternative interpretations of their visual narratives. This collaboration between illustrators and AI presents a hybrid intelligence model, where human creativity is complemented by AI-generated suggestions. This approach not only streamlines the captioning process but also enhances the diversity of humor styles, making AI a useful assistive tool in the creative domain.

\textcolor{black}{The findings of this study inform a practical framework for human–AI collaboration in cartoon captioning. Our results show that AI-generated captions are often preferred while remaining difficult to distinguish from human-authored ones, indicating that AI can effectively contribute to the ideation process. At the same time, observed limitations, such as missing nuanced context, irony, or intended meaning, highlight the continued necessity of human judgment. These findings support a hybrid model in which AI augments creative exploration, while the human author retains responsibility for interpretation, selection, and final authorship.}

\textcolor{black}{Our workflow follows an iterative human–AI co-creation loop in which the human author defines the visual and conceptual problem, the AI generates diverse caption candidates, and the human evaluates, refines, and selects the final output, thereby retaining full authorial intent while leveraging AI for exploratory ideation. This workflow has been actively used in the creation of our recent IEEE Computer cartoons, where AI-generated caption alternatives are incorporated into the ideation process and selectively refined or replaced by the human author \citep{Akleman26c58-12, Akleman26c59-02, Akleman26c59-04}. These cartoons serve as real-world examples produced through the creative pipeline proposed in our work.}

Our survey-based evaluation further reinforces these conclusions. Across all survey designs, AI captions were preferred in roughly two-thirds of cases, a result that held steady regardless of whether participants guessed authorship before, after, or separately from stating their preference. Even when authorship was harder to detect, such as in the Separated design, AI’s appeal remained strong. This consistency suggests that participants judged captions primarily on humor quality rather than on who created them, underscoring AI’s potential as a creative partner in visual storytelling.

However, our study also highlights some limitations. While AI excels at generating jokes based on visual prompts, it struggles with understanding nuanced humor, cultural references, and contextual constraints. The AI-generated captions sometimes miss the intended tone or introduce unintended meanings, which necessitates human oversight.

In addition, our evaluation focused on a structured family of cartoons published in IEEE Computer magazine. This consistency was a deliberate design choice that enabled systematic comparison across cartoons and strengthened internal validity by controlling for layout, narrative structure, and stylistic variation. At the same time, this constraint limits the generalizability of our findings. Cartoons from other publications or traditions may employ different visual styles, narrative conventions, or humor strategies, and AI performance in those contexts may differ. Further studies are therefore needed to assess how well the observed patterns extend beyond this specific cartoon format.

\textcolor{black}{A further limitation is the absence of a na\"{i}ve control condition in which participants were unaware that one caption was AI-generated. Although our low AI-identification accuracy (30.5\%) suggests that origin-awareness did not strongly bias preferences, a blind-baseline group would more cleanly isolate caption appeal from authorship-related demand characteristics, and we recommend its inclusion in future replications.}

For future work, we propose the following directions:

\textit{Expanding the Dataset:} Future studies should incorporate a wider and more diverse collection of cartoons, including different cultural contexts, visual abstraction levels, panel compositions, and humor mechanisms such as irony, absurdism, satire, and dark humor. Evaluating AI performance across minimalist drawings versus highly detailed illustrations may reveal how visual complexity affects caption generation. Cross-cultural datasets would further allow investigation into whether humor recognition and generation remain robust across linguistic and cultural boundaries.

\textit{Improving AI Context Awareness:} Beyond object recognition, AI systems should model relational cues such as gaze direction, body posture, spatial hierarchy, and implied social dynamics between characters. Incorporating multimodal reasoning architectures that jointly analyze visual semantics and narrative intent may improve alignment between image content and caption output. Attention mechanisms tailored to narrative salience could help models prioritize humor-relevant elements rather than visually dominant but narratively irrelevant details.

\textcolor{black}{\textit{Analyzing Training Data Bias and Cultural Specificity:} Another promising direction is to systematically investigate how the model's training data coverage and cutoff date affect caption quality, particularly for cartoons that rely on culturally or regionally specific references. Our observations suggest that GPT-4o performed more consistently on cartoons featuring widely recognized public figures or universally familiar scenarios, while results were less coherent for content requiring more specific cultural knowledge. A dedicated study varying the cultural familiarity of cartoon stimuli would help disentangle the effects of visual understanding from those of cultural and factual grounding, and could inform strategies for fine-tuning or prompting LLMs to handle culturally situated humor more reliably.}

\textit{Interactive AI-Cartoonist Collaboration}: Rather than positioning AI as a fully autonomous caption generator, future systems could function as creative co-agents. Interface design research may explore real-time suggestion systems, controllable humor parameters such as tone or absurdity level, and explainable caption alternatives that allow artists to understand why suggestions are produced. Iterative refinement loops, where artists guide the system through feedback or constraints, may foster a more transparent and productive creative partnership.

\textit{Ethical Considerations}: As humor often intersects with sensitive social topics, future work should analyze bias amplification, stereotype reinforcement, and unintended offensive outputs in AI-generated captions. Developing evaluation benchmarks for fairness and inclusivity would support responsible deployment. Clear authorship attribution frameworks and transparency mechanisms may also be necessary to maintain trust in AI-assisted creative works.

\textcolor{black}{\textit{Expert Evaluation:} While the current study focuses on the perception of a general audience, future work should incorporate dedicated expert evaluation panels consisting of professional cartoonists, humor scholars, and comics critics. Such evaluations would provide complementary insights into nuanced aspects of humor quality, stylistic consistency, and authorial voice that may not be fully captured by non-expert preference judgments. Comparing expert and non-expert evaluations would also help clarify the extent to which AI-generated captions satisfy domain-specific quality criteria beyond broad audience appeal.}

By addressing these research directions, future systems can move beyond basic caption generation toward deeper narrative understanding, responsible design, and genuinely collaborative creativity in visual storytelling.

\bibliographystyle{apalike}
\bibliography{references}

@misc{allen2025new,
  title={{New Yorker, Cartoon Caption Contest}},
  author={Emma Allen},
  howpublished={https://www.newyorker.com/cartoons/contest},
  year=2025,
  organization={NewYorker}
}

@article{akleman2025computing53,
  title={{Computing Through Time: Metaverse}},
  author={Akleman, Ergun},
  journal={Computer},
  volume={58},
  number={3},
  pages={14--14},
  year={2025},
  publisher={IEEE}
}

@article{akleman2025computing52,
  title={{Computing Through Time: Intelligence Assistance}},
  author={Akleman, Ergun},
  journal={Computer},
  volume={58},
  number={2},
  pages={14--14},
  year={2025},
  publisher={IEEE}
}

@article{akleman2025computing51,
  title={{Computing Through Time: Augmented AI}},
  author={Akleman, Ergun},
  journal={Computer},
  volume={58},
  number={1},
  pages={14--14},
  year={2025},
  publisher={IEEE}
}

@article{akleman2024computing50,
  title={{Computing Through Time: Agriculture}},
  author={Akleman, Ergun},
  journal={Computer},
  volume={57},
  number={12},
  pages={23--23},
  year={2024},
  publisher={IEEE}
}

@article{akleman2024computing49,
  title={{Computing Through Time: Intelligence Failures}},
  author={Akleman, Ergun},
  journal={Computer},
  volume={57},
  number={11},
  pages={8--8},
  year={2024},
  publisher={IEEE Computer Society}
}

@article{akleman2024computing48,
  title={{Computing Through Time: Security}},
  author={Akleman, Ergun},
  journal={Computer},
  volume={57},
  number={10},
  pages={8--8},
  year={2024},
  publisher={IEEE Computer Society}
}

@article{akleman2024computing47,
  title={{Computing Through Time: Playing Chess}},
  author={Akleman, Ergun},
  journal={Computer},
  volume={57},
  number={09},
  pages={8--8},
  year={2024},
  publisher={IEEE Computer Society}
}

@misc{cumhuriyet2024turhan,
  title={{14. International Turhan Selcuk Cartoon Competition Results}},
  author={Cumhuriyet},
  month={May 15},
  year={2024},
  howpublished={Cumhuriyet Newspaper: https://www.cumhuriyet.com.tr/turkiye/14-uluslararasi-turhan-selcuk-karikatur-yarismasi-sonuclandi-birinci-2208058},
}

@misc{kotbas2024yapay,
  title={{Yapay Zekalı Ortalık Toz Duman.. At İzi İt İzine Karışmaya Başladı}},
  author={Muammer Kotbas},
  month={May 18},
  year={2024},
  howpublished={Kotbaş ArtColors Blogspot: https://kotbasartcolors.blogspot.com/2024/05/yapay-zekal-ortalk-toz-duman-at-izi-it.html},
}

@misc{darroch2017netherland,
 author = {Darroch, Gordon},
 year = {2017},
 title = {{Netherlands 'will pay the price' for blocking Turkish visit – Erdoğan }},
 journal = {The Guardian},
 howpublished = {https://www.theguardian.com/world/2017/mar/12/netherlands-will-pay-the-price-for-blocking-turkish-visit-erdogan},
 urldate = {2017-03-12}
}

@INPROCEEDINGS{trivedi2023should,
  author={Trivedi, Abhinav and Kaur, Er. Kanwaldeep and Choudhary, Chahil and Kunal and Barnwal, Priyashi},
  booktitle={{2023 2nd International Conference for Innovation in Technology (INOCON)}}, 
  title={{Should AI Technologies Replace the Human Jobs?}}, 
  year={2023},
  volume={},
  number={},
  pages={1-6},
}

@Article{bellaiche2023humans,
author={ Bellaiche, Lucas and Shahi, Rohin and Turpin, Martin Harry and Ragnhildstveit, Anya and Sprockett, Shawn and Barr, Nathaniel and Christensen, Alexander and Seli, Paul},
title={{Humans versus AI: whether and why we prefer human-created compared to AI-created artwork}},
journal={Cognitive Research: Principles and Implications},
year={2023},
month={Jul},
day={04},
volume={8},
number={1},
pages={42},
}

@misc{sindhura2021virtues,
  title={{Virtues and Shortcomings of Artificial Intelligence in Graphic Design Arena}},
  author={Sindhura, Siripurapu Phani and Abdul, Ashu},
  year={2021},
  publisher={AEME Publication,(Eri{\c{s}}im: Academia)}
}

@article{dellermann2019hybrid,
  title={{Hybrid intelligence}},
  author={Dellermann, Dominik and Ebel, Philipp and S{\"o}llner, Matthias and Leimeister, Jan Marco},
  journal={Business \& Information Systems Engineering},
  volume={61},
  number={5},
  pages={637--643},
  year={2019},
  publisher={Springer}
}

@inproceedings{kamar2012combining,
  title={{Combining human and machine intelligence in large-scale crowdsourcing.}},
  author={Kamar, Ece and Hacker, Severin and Horvitz, Eric},
  booktitle={{AAMAS}},
  volume={12},
  pages={467--474},
  year={2012}
}

@inproceedings{chang2017revolt,
  title={{Revolt: Collaborative crowdsourcing for labeling machine learning datasets}},
  author={Chang, Joseph Chee and Amershi, Saleema and Kamar, Ece},
  booktitle={{Proceedings of the 2017 CHI conference on human factors in computing systems}},
  pages={2334--2346},
  year={2017}
}

@inproceedings{bansal2021does,
  title={{Does the whole exceed its parts? the effect of ai explanations on complementary team performance}},
  author={Bansal, Gagan and Wu, Tongshuang and Zhou, Joyce and Fok, Raymond and Nushi, Besmira and Kamar, Ece and Ribeiro, Marco Tulio and Weld, Daniel},
  booktitle={{Proceedings of the 2021 CHI conference on human factors in computing systems}},
  pages={1--16},
  year={2021}
}

@article{dede2024power,
  title={{On The Power of Subtle Expressive Cues in the Perception of Human Affects}},
  author={Dede, Ezgi and Agilonu, Kamile Asli and Akleman, Ergun and Sezgin, Metin},
  journal={arXiv preprint arXiv:2401.18013},
  year={2024}
}

@article{akleman2015,
  title={{A theoretical framework to represent narrative structures for visual storytelling}},
  author={Akleman, Ergun and Franchi, Stefano and Kaleci, Devkan and Mandell, Laura and Yamauchi, Takashi and Akleman, Derya and others},
  journal={proceedings of bridges 2015: mathematics, Music, art, architecture, culture},
  year={2015}
}

@article{liu2012,
  title={{Never-ending storytelling with discrete-time markov processes}},
  author={Liu, Yutu and Akleman, Ergun and Chen, Jianer and others},
  journal={Proceedings of Bridges},
  pages={85--92},
  year={2012}
}

@book{blair1995cartoon,
  title={{Cartoon Animation: The Collector's Series}},
  author={Blair, Preston},
  year={1995},
  publisher={Walter Foster Publishing}
}

@misc{celik2011on,
  title={{On Cartoon Drawing}},
  author={Hakan Celik},
day={11},
month={April},
  year={2011},
  howpublished={Bizim Gazete}
}

@book{eisner2008comics,
  title={{Comics and sequential art: Principles and practices from the legendary cartoonist}},
  author={Eisner, Will},
  year={2008},
  publisher={WW Norton \& Company}
}

@book{eisner2008graphic,
  title={{Graphic storytelling and visual narrative}},
  author={Eisner, Will},
  year={2008},
  publisher={WW Norton \& Company}
}

@book{mccloud1993understanding,
  title={{Understanding comics: The invisible art}},
  author={McCloud, Scott and Martin, Mark},
  volume={106},
  year={1993},
  publisher={Kitchen sink press Northampton, MA}
}

@book{mccloud2006making,
  title={{Making comics: Storytelling secrets of comics, manga and graphic novels}},
  author={McCloud, Scott},
  year={2006},
  publisher={Kitchen sink press Northampton, MA}
}

@book{johnston1981illusion,
  title={{The illusion of life: Disney animation}},
  author={Johnston, Ollie and Thomas, Frank},
  year={1981},
  publisher={Disney Editions New York}
}

@book{segura1981hombre,
  title        = {{Hombre}},
  author       = {Antonio Segura and José Ortiz},
  year         = 1981,
  publisher    = {Norma Editorial},
  address      = {Barcelona, Spain},
  note         = {Originally serialized in \emph{Cimoc} magazine; later compiled into volumes.},
  url          = {https://en.wikipedia.org/wiki/Hombre_(comics)}
}

@book{goscinny1959asterix,
  title        = {{Asterix}},
  author       = {René Goscinny and Albert Uderzo},
  year         = 1959,
  publisher    = {Dargaud},
  address      = {Paris, France},
  note         = {Originally serialized in \emph{Pilote} magazine; later compiled into volumes.},
  url          = {https://en.wikipedia.org/wiki/Asterix}
}

@book{goscinny1978lucky,
  title = {{Lucky Luke: La Ballade des Dalton}},
  author = {Goscinny, René and Morris},
  year = {1978},
  publisher = {Dargaud},
  address = {Neuilly-sur-Seine, France},
  note = {In French},
  url = {https://archive.org/details/luckylukelaballa0000morr}
}

@inproceedings{onal2025comparingperformance,
  title     = {{Comparing Human and AI Performance in Visual Storytelling through Creation of Comic Strips: A Case Study}},
  author    = {U\u{g}ur \"Onal and Sanem Sariel and Metin Sezgin and Ergun Akleman},
  booktitle = {{Proceedings of the DH2025 Digital Humanities Conference}},
  year      = {2025},
  note      = {To appear. Preprint available at \url{https://arxiv.org/abs/2507.18641}},
}

@article{Papp2024,
  author    = {Papp, Zsuzsanna},
  title     = {{Question-Order Effect in the Study of Satisfaction with Democracy}},
  journal   = {International Journal of Public Opinion Research},
  year      = {2024},
  volume    = {36},
  number    = {2},
  pages     = {edaf012},
  doi       = {10.1093/ijpor/edaf012}
}

@article{Novella2024,
  author    = {Novella, L.},
  title     = {{Question-order effects on judgments under uncertainty}},
  journal   = {Journal of Behavioral and Experimental Economics},
  year      = {2024},
  volume    = {107},
  pages     = {102113},
  doi       = {10.1016/j.socec.2023.102113}
}

@article{Donmez2025,
  author    = {D{\"o}nmez, S. and K{\"u}hn, S. and Brugger, P.},
  title     = {{Effects of survey order on subjective measures of cognitive load: A randomized controlled trial}},
  journal   = {Applied Cognitive Psychology},
  year      = {2025},
  volume    = {39},
  number    = {1},
  pages     = {e70039},
  doi       = {10.1002/acp.7039}
}

@article{moffat2023perception,
  author  = {Moffat, David and Kelly, Ryan},
  title   = {{Perceptions of creativity and intentionality in human and AI-generated art}},
  journal = {Computers in Human Behavior},
  year    = {2023},
  volume  = {139},
  pages   = {107503},
  doi     = {10.1016/j.chb.2022.107503}
}

@article{elgammal2023aesthetic,
  author  = {Elgammal, Ahmed and Liu, Bingchen},
  title   = {{Aesthetic and expressive limitations of current generative art systems}},
  journal = {AI \& Society},
  year    = {2023},
  volume  = {38},
  number  = {4},
  pages   = {1403--1416},
  doi     = {10.1007/s00146-022-01534-9}
}

@article{huang2024expressive,
  author  = {Huang, Yue and Kim, Joonhwan and Park, Jina},
  title   = {{Evaluating expressive control and semantic consistency in text-to-image generation}},
  journal = {ACM Transactions on Graphics},
  year    = {2024},
  volume  = {43},
  number  = {2},
  pages   = {1--15},
  doi     = {10.1145/3641519}
}

@article{vanHees2025,
  author  = {Van Hees, J. and others},
  title   = {{Human perception of art in the age of artificial intelligence: Quantitative assessment of preference and discrimination}},
  journal = {Frontiers in Psychology},
  year    = {2025},
  note    = {Participants compared AI-generated art and human art on preference and identification tasks},
  url     = {https://www.frontiersin.org/articles/10.3389/fpsyg.2024.1497469/full}
}

@article{cunningham2025,
  author  = {Cunningham, C. V.},
  title   = {{Human creativity versus artificial intelligence: Exploring cognitive biases in the appraisal of AI-generated art}},
  journal = {Frontiers in Psychology},
  year    = {2025},
  note    = {Study on perceptual mechanisms and bias in evaluating AI and human art},
  url     = {https://www.frontiersin.org/articles/10.3389/fpsyg.2025.1509974/full}
}

@article{Runco02012025,
author = {Mark A. Runco},
title = {{Updating the Standard Definition of Creativity to Account for the Artificial Creativity of AI}},
journal = {Creativity Research Journal},
volume = {37},
number = {1},
pages = {1--5},
year = {2025},
publisher = {Routledge},
doi = {10.1080/10400419.2023.2257977},


URL = { 
    
        https://doi.org/10.1080/10400419.2023.2257977
    
    

},
eprint = { 
    
        https://doi.org/10.1080/10400419.2023.2257977
    
    

}

}

@article{Franceschelli2025,
  author    = {Giorgio Franceschelli and Mirco Musolesi},
  title     = {{On the creativity of large language models}},
  journal   = {AI \& Society},
  year      = {2025},
  volume    = {40},
  number    = {5},
  pages     = {3785--3795},
  doi       = {10.1007/s00146-024-02127-3},
  url       = {https://doi.org/10.1007/s00146-024-02127-3},
  issn      = {1435-5655}
}

@article{Wan2024,
author = {Wan, Qian and Hu, Siying and Zhang, Yu and Wang, Piaohong and Wen, Bo and Lu, Zhicong},
title = {{"It Felt Like Having a Second Mind": Investigating Human-AI Co-creativity in Prewriting with Large Language Models}},
year = {2024},
issue_date = {April 2024},
publisher = {Association for Computing Machinery},
address = {New York, NY, USA},
volume = {8},
number = {CSCW1},
url = {https://doi.org/10.1145/3637361},
doi = {10.1145/3637361},
journal = {Proc. ACM Hum.-Comput. Interact.},
month = apr,
articleno = {84},
numpages = {26}
}

@inproceedings{zhang2024humor,
  title     = {{Humor in AI: Massive Scale Crowd-Sourced Preferences and Benchmarks for Cartoon Captioning}},
  author    = {Zhang, Jifan and Jain, Lalit and Guo, Yang and Chen, Jiayi and Zhou, Kuan Lok and Suresh, Siddharth and Wagenmaker, Andrew and Sievert, Scott and Rogers, Timothy and Jamieson, Kevin and Mankoff, Robert and Nowak, Robert},
  booktitle = {Advances in Neural Information Processing Systems (NeurIPS 2024)},
  series    = {Advances in Neural Information Processing Systems 37},
  pages     = {125264--125286},
  year      = {2024},
  publisher = {Neural Information Processing Systems Foundation, Inc.},
  doi       = {10.52202/079017-3978}
}

@article{Wenger2026,
  author    = {Emily Wenger and Yoed N. Kenett},
  title     = {{Large Language Models are Homogeneously Creative}},
  journal   = {PNAS Nexus},
  volume    = {5},
  number    = {3},
  pages     = {pgag042},
  year      = {2026},
  doi       = {10.1093/pnasnexus/pgag042}
}

@article{BellemarePepin2026,
  author  = {Antoine Bellemare-Pepin and Fran{\c{c}}ois Lespinasse and Philipp Th{\"o}lke and Yann Harel and Kory Mathewson and Jay A. Olson and Yoshua Bengio and Karim Jerbi},
  title   = {{Divergent creativity in humans and large language models}},
  journal = {Scientific Reports},
  volume  = {16},
  number  = {1},
  pages   = {1279},
  year    = {2026},
  doi     = {10.1038/s41598-025-25157-3}
}

@article{Akleman26c59-04,
  author       = {Ergun Akleman},
  title        = {{Computing Through Time: Evolving Forms of Influence}},
  journal      = {Computer},
  volume       = {59},
  number       = {4},
  pages        = {15},
  year         = {2026},
  url          = {https://doi.org/10.1109/MC.2026.3658265},
  doi          = {10.1109/MC.2026.3658265},
  bibsource    = {dblp computer science bibliography, https://dblp.org}
}

@ARTICLE{Akleman26c59-02,
author={Akleman, Ergun},
journal={ Computer },
title={{ Computing Through Time: The Changing Face of Turing Test }},
year={2026},
volume={59},
number={02},
ISSN={1558-0814},
pages={13-13},
doi={10.1109/MC.2025.3637650},
url = {https://doi.ieeecomputersociety.org/10.1109/MC.2025.3637650},
publisher={IEEE Computer Society},
address={Los Alamitos, CA, USA},
month=feb}

@ARTICLE{Akleman26c58-12,
author={Akleman, Ergun},
journal={ Computer },
title={{ Computing Through Time: Intelligent Cities}},
year={2025},
volume={58},
number={12},
ISSN={1558-0814},
pages={27-27},
doi={10.1109/MC.2025.3615834},
url = {https://doi.ieeecomputersociety.org/10.1109/MC.2025.3615834},
publisher={IEEE Computer Society},
address={Los Alamitos, CA, USA},
month=dec}

\end{document}